# Energy Requirement of Control: Comments on Szilard's Engine and Maxwell's Demon

L. B. KISH [(1)] , C.G. GRANQVIST [(2)]

[(1)] *Department of Electrical and Computer Engineering, Texas A&M University - College Station, TX 77843-3128, USA*
[(2)] *Department of Engineering Sciences, The Ångström Laboratory, Uppsala University, P. O. Box 534, SE-75121 Uppsala, Sweden*

**Abstract** – In mathematical physical analyses of Szilard's engine and Maxwell's demon, a general assumption (explicit or implicit) is that one can neglect the energy needed for relocating the piston in Szilard's engine and for driving the trap door in Maxwell's demon. If this basic assumption is wrong, then the conclusions of a vast literature on the implications of the Second Law of Thermodynamics and of Landauer's erasure theorem are incorrect too. Our analyses of the fundamental information physical aspects of various type of control within Szilard's engine and Maxwell's demon indicate that the entropy production due to the necessary generation of information yield much greater energy dissipation than the energy Szilard's engine is able to produce even if all sources of dissipation in the rest of these demons (due to measurement, decision, memory, etc) are neglected.

## 1. Introduction: Demons and the need of control

– Heat engines (e.g., [1]) use a temperature difference to produce work while heat demons (information demons), such as Szilard's engine [2] and Maxwell's demon [3], employ information on the instantaneous amplitude of thermal fluctuations and execute active control to produce a temperature difference or work. A useful collection of papers on this topic was presented recently [4]. In particular we want to draw attention to seminal work by Brillouin [5,6]. Our present paper raises the question of energy requirements for active and passive control, which to our knowledge has not been done before. The lack of such analyses makes it is understandable that existing papers regard (implicitly or explicitly) the relocation of the piston of Szilard's engine and the control of the trap door of Maxwell's demon as operations that ideally do not demand energy, or that they need less energy than the demon is supposed to gain. In the opposite case—i.e., if an ultimate energy requirement exists which is beyond the energy that an ideal demon can produce—many earlier conclusions related to the Second Law of Thermodynamics [4] and the claimed necessity of Landauer's erasure theorem [4, 7-12] in the analysis of Szilard's engine need reassessment. For example, if these control operations themselves demand more energy than Szilard's engine can produce, then the Second Law of Thermodynamics is *not* violated and Landauer's erasure theorem is *not* needed in order to restore it.

Perhaps, the main reason for the lack of studies of the kind referred to above is that these historical demons [2,3] basically are mechanical systems coupled to a heat reservoir where different types of energies coexist. In addition to the obvious heat and mechanical energies in these systems, energies (often of different type, such as electrical or photonic) are needed for the information collection (measurements, monitoring), decision (logic operations) and control. Information processing, decision, control, and system analysis typically are part of electronics, where these concepts mostly have been developed, and they do not belong to conventional physics; this may explain why they have been neglected. In the present paper we focus on the energy requirement for active and passive control and show that they have fundamental minimum values that are inherently related to the errors of the operation and that this energy, in the case of Szilard's engine, is beyond the energy produced by the ideal engine. Thus the energy needed for active control will itself be enough to avoid the violation of the Second Law of Thermodynamics so that Landauer's erasure theorem is not needed.

## 2. Types of controlled actions/motions in the demons

– For determining the necessary control information, the simple rule of thumb is that a yes/no or on/off type of control is one-bit information if the control is error-free.

For example information related to the question whether the thermal velocity of the piston in Szilard's engine points in the proper direction is one-bit information. When not only the sign but also the actual value of the velocity is needed, this represents extra bits of information determined by the required accuracy.

To run Szilard's engine we shall control the velocity of the piston when, at the end of the cycle, the piston is relocated from the end of the cylinder to its middle. The process contains five stages, as discussed below, and the motion itself contains three stages as shown in Figure 1. The disengagement from the gearbox is essential; otherwise we must invest the same energy in the relocation of the piston as the work the engine performed during the expansion. In the list of the five stages, the amount of minimum control information is also indicated:

(0) When the piston reaches the end of the cylinder, disengaging the clutch that couples the piston to the gearbox: 1 bit

(1) Starting the motion: checking if the thermal velocity of the piston points in the correct direction; otherwise injecting energy/momentum to reverse it: 1 bit (if it is in the correct direction, more if not)

(2) Continuous motion to the desired position: >1 bit

(3) Stopping the motion when the piston reaches its final position: 1 bit

(4) Reengaging the clutch to couple the piston to the gearbox: 1 bit

One should note that the speed in stage (2) and the duration of the relocation are unimportant, but the moment of arrival at the final position must be monitored in order to prepare for stage (3).

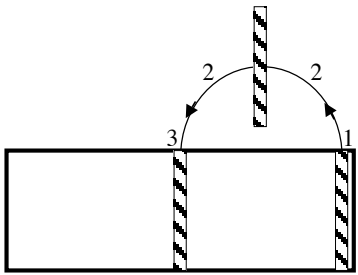

Fig 1. Stages of motion requiring different control actions in Szilard's engine: starting (1), continuous motion (2), stopping (3).

The trapdoor in Maxwell's demon is somewhat different and must be controlled in a timed fashion in order to pass the arriving "hot" molecules and reflect the "cold" ones. The process contains eight stages, and the motion itself has six stages as illustrated in Figure 2. The existence and the disengagement/engagement of the lock at the closed trapdoor position are essential; otherwise the trapdoor would diffuse away from its location. The eight stages are given below, and the minimum required control information is indicated as well:

(0) Disengaging the lock holding the trapdoor in position: 1 bit

(1) Starting the motion of trapdoor opening: providing energy/momentum. The velocity must overcome the actual thermal velocity-fluctuation-amplitude and must have a required relative accuracy for proper timing of door opening: ≥ 1 bit

(2) Continuous motion until the door opens sufficiently. Either measuring the status of the door or the time: ≥ 1 bit

(3) Letting the molecule pass at open trapdoor (which may still move). Control information included at (1)

(4) Reversing the motion to close the trapdoor: 1 bit

(5) Continuous motion until the door reaches its closed position and detecting that: ≥ 1 bit

(6) Stopping the motion when the door is totally closed: 1 bit

(7) Reengaging the lock holding the trapdoor in position: 1 bit

The various stages delineated above require a few remarks: Stage (2) of Szilard's engine and stage (5) of Maxwell's demon need monitoring, which involves several measurements with yes/no answers (to questions such as "has the body arrived at its required position?"). However the seemingly similar stage (2) of Maxwell's demon demands another type of control information based on time measurement and its relative inaccuracy. At stage (1) a velocity of proper accuracy—in order to give the timing for the total opening when the molecule arrives—has been provided, and thus the demon can utilize time measurement instead of position monitoring. Nevertheless both cases require control information of at least 1 bit (one on/off type operation, or more).

In conclusion, the generated control information $I_{Sz}$ of Szilard's engine and $I_{Ma}$ of Maxwell's demon can be written

$$I_{Sz} > 5 \text{ bits} \tag{1}$$

and

$$I_{Ma} > 7 \text{ bits.} \tag{2}$$

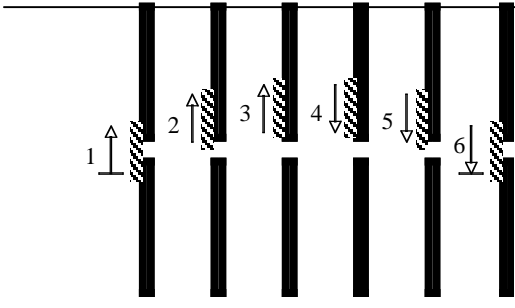

Fig. 2. Stages of motion requiring different control actions in Maxwell's demon: starting (1), continuum motion (2,3), reversal of motion (4), continuum motion (5), stopping (6).

## 3. Fundamental energy dissipation limits for control

– We now deduce fundamental lower limits of the energy dissipation for control in Szilard's engine and Maxwell's demon. These fundamental lower limits, which are valid for theoretically ideal systems, are calculated from Brillouin's *negentropy principle of information* [5,6]. It is important to note that, in physical situations, one can give values larger than those dictated for the lower limits of energy dissipation by Brillouin's negentropy principle; these issues are further discussed in Section 4. Nevertheless, the negentropy principle yields a solid lower limit that is comparatively easy to compute, and the results are conclusive for our goals. Thus we explore this direction first.

According to Brillouin's negentropy principle of information [5,6], setting up a signal or obtaining a measurement result at a classical physical output with information content of $I_s$ bits requires the emission of at least an amount of entropy $dS_s$ to the environment according to

$$dS_s = k I_s \ln(2) ,\qquad(3)$$

where $k$ is Boltzmann's constant and the term $\ln(2)$ originates from the specifically used unit (i.e., the bit) of information [5]. This entropy is equivalent to the emission of a corresponding minimum heat $Q_s$ required to generate the signal, viz.

$$Q_s = T dS_s = k T I_s \ln(2) ,\qquad(4)$$

where $T$ is the absolute temperature.

Relations (1), (2) and (4) yield the energy dissipation caused by controlling information in Szilard's engine and Maxwell's demon by

$$Q_{Sz} = k T I_{Sz} \ln(2) > 5 k T \ln(2) \approx 3.5 k T \qquad(5)$$

and

$$Q_{Ma} = k T I_{Ma} \ln(2) > 7 k T \ln(2) \approx 5 k T .\qquad(6)$$

In Szilard's engine, the lower limit for the energy requirement of control is five times larger than the useful work the engine is able to produce, i.e., $kT \ln(2)$ [2]. Thus the need for control, which is an essential part of Szilard's engine, itself secures that the Second Law of Thermodynamics is not violated.

The case of Maxwell's demon is different because—even though its energy requirement for control is greater—one can set the energy selection rules so that the energy $E_{mol}$ of the chosen "hot" molecule is much greater than the dissipated energy described above. However it is easy to resolve this paradox, and the demon must measure and generate new information about all incoming molecules albeit it can produce energy only from an exponentially small fraction $\kappa$ of molecules scaling with the Boltzmann factor, where

$$\kappa \propto \exp\left(-\frac{E_{mol}}{kT}\right) << 1 .\qquad(7)$$

Thus even though Maxwell's demon will let through a very small fraction $\kappa$ of incoming molecules and generate control-related dissipation, all of the other molecules will contribute to measurement-related new information. Consequently the measurement process itself will save the Second Law of Thermodynamics, as has been pointed out before [2,5,6].

## 4. Real ideal limits versus limits dictated by Brillouin's negentropy

– Recent results [13,14] exploring level-crossing statistics of thermal fluctuations and their implications for energy dissipation make it possible to test the meaning of Brillouin's negentropy limits in a control system. To this end we first analyze the case of binary control, which includes a two-stage device (i.e., a switch). In particular, the information request/serving process always involves a full (open/close) cycling of a switch for each information bit. To stabilize the actual state of such a device and to reduce errors, an energy threshold $E_{th}$ separates the two stages. Performing a full cycle with the switch leads to an energy dissipation of at least $E_{th}$ [13,14]. (Switches with a double-well-potential scheme and $E_{th}$ energy barrier dissipate an energy of $2E_{th}$ during a full cycle, while switches with an asymmetric potential energy scheme and $E_{th}$ energy difference invest $E_{th}$ in one half-cycle and dissipate the same energy in the other half-cycle [13,14].)

The specific value of $E_{th}$ is crucial for the error-probability of operation. Thermal fluctuations are Gaussian noise processes, and they will cross any finite-energy threshold during a sufficiently long waiting time. The minimum threshold energy $E_{th,min}$ and corresponding energy dissipation to carry out a binary single-bit cycling operation was analyzed recently [14] by the threshold-crossing statistics of Gaussian processes (due to the expanded Rice-formula [13]) and the result is

$$E_{th,min} = kT \ln\left(\frac{2}{\sqrt{3}} \frac{1}{\varepsilon}\right) \approx kT \ln\left(\frac{1}{\varepsilon}\right) \ , \qquad (8)$$

where $\varepsilon$ is the bit error probability during the observation time $\tau_{min}$; it is equal to the reciprocal bandwidth, which is also the time-resolution limit in the system. Here $\varepsilon = 0.5$ represents the case of zero information [15] (for example throwing a random coin) and, equivalently, zero efficiency when this single-bit information is used for controlling an engine. Furthermore $\tau_{min}$ is the autocorrelation time of the thermal fluctuations in the system [14], and hence the errors generated in non-overlapping $\tau_{min}$ intervals are independent. Thus, in accordance with Equation (8), long observation times with $\tau_{obs} >> \tau_{min}$ require $E_{th,min}$ to grow logarithmically with increasing $\tau_{obs} / \tau_{min}$ ratio in order to keep the same $\varepsilon$, thus reiterating that Brillouin's negentropy approach gives lower estimations than the actual, more strict lower limits of energy dissipation.

It is interesting to note that some of the control stages described in Section 2 involve more than a single bit of information. The natural question is then whether it is better to use analog than binary information in those cases. The Shannon–Hartley formula [5,6],

$$I_{an} = 0.5 \log_2\left(1 + \frac{E_{sig}}{E_{flu}}\right) , \qquad (9)$$

describes the maximum information content $I_{an}$ of an analog channel at any given time in terms of the mean-square energy $E_{sig}$ of the signal and that of $E_{flu}$ for the random fluctuations. When the analog channel is a physical system where the noise is the thermal fluctuation (thermal noise), it is obvious from equation (9) that such systems are worse concerning energy dissipation for simultaneous multi-bit operations than parallel binary channels representing the same amount of (multi-bit) information. According to equation (9), the information content scales in the analog case with the logarithmic function of the invested signal energy, which means that the energy dissipation is an exponential function of the information content.

Summarizing, the simple results shown in Section 3 provide valid lower limits and give decisive answers for the ultimate question in this paper. However the lower limits of energy dissipation deduced from Brillouin's negentropy are extremely optimistic and represent a binary case where the dissipation is lowest, the possible error probability is highest, and the engines have zero efficiency. For a real, functioning engine, however, the error probability must be decreased and the energy thresholds and dissipation correspondingly increased.

Furthermore, if the control process requires longer times than the reciprocal bandwidth of the system—as it naturally does—there is a further logarithmic growth factor in the energy dissipation to keep the error probability $\varepsilon$ at the required low level and the engine's efficiently at the required non-zero level.

Finally, it should be noted that all of the above considerations were given for lossless systems with ideal elements and transmitters. The real energy dissipation of any physical information demon will always be greater than the ideal values for reasons related to technical limitations and non-idealities such as non-zero losses and friction and efficiencies of transducers etc being less than 100%.

**5. Related topics** – Finally in this section and in the next one, we take a look at the bigger picture and briefly discuss some of the topics related to the energy dissipation problem analyzed above: Landauer's theorem and reversible computation; energy dissipation in quantum computing; the often omitted essential component of addressing the energy dissipation issue: addressing the error probability; and some recent experiments.

There has been an ongoing discussion for years about the validity of Landauer's erasure theorem [7-12]. The theorem claims that in information processing the erasure is the process that is fundamentally dissipative whereas dissipation does not take place for creating or writing new information. Recently, Norton has published two extensive studies [7,8] about problems with this theorem; a brief survey of his results and arguments, and a refusal of energy-free computing, were presented recently [9]. For example, Norton points out that theories aimed at making a fundamental proof of Landauer's theorem "selectively neglect thermal fluctuations" [7]. Concerning the claim of energy-dissipation-free measurements, he states that "concrete proposals for how we might measure dissipationlessly and expand single molecule gases reversibly are shown to be fatally disrupted by fluctuations" [7].

All these discussions belong to a larger set of debates about the ultimate energy requirement of information processing and of energy-free (thermodynamically reversible) computing [9,10-16]. For example, in a refusal of the reversible computing initiative it was pointed out [11,12] that logical reversibility is possible but that this fact in itself does not imply physical reversibility, the decisive underlying reason being thermal fluctuations [9].

According to Equation (8), no logic operation is reversible in a classical physical computer circuit. But even the hopes that quantum informatics can offer lower energy dissipation vanished when it was demonstrated [16,17] that an ideal general-purpose quantum computer would dissipate three orders of magnitude more heat than an ideal classical computer with the same information processing and calculation performance. The reason is the ultimate uncertainty/reversibility of gates due to the energy-time uncertainly principle [18] that results in not only the need of error corrections but also situations when error correction is impossible [19,20].

We reiterate that any theory or analysis that neglects errors and does not contain them specifically in the equations of energy balance (either in the form of entropy or error probability) and/or of the fluctuations (thermal or quantum) causing them has very limited value and any predicted minimum energy requirement is incorrect. Specifically, as seen in Equation (8), zero error probability implies infinite energy demand for an elementary logic operation in a classical physical logic circuit. Thus supposing that no fluctuations are present and still arriving at finite energy dissipation means that an infinite energy dissipation term is ignored.

**6. A few words about recent experiments** – Finally, we briefly address recent experimental works on specific classical physical systems [21, 22] where a particle in a thermal bath is controlled by properly controlling its potential.

In [21], it was demonstrated that the particle was able to climb up on the potential-ladder, and the authors' goal was to confirm the Jarzynski equality [23] in the context of utilizing information on thermal fluctuations to turn them into work. This does not mean that information demons have been built and demonstrated, though! Specifically, in the above example [21] the fact that the particle climbed up the ladder proved that an *engine* consuming external energy and doing work, and not an information demon utilizing control to extract energy from thermal motion, was built and working. The fact that the mentioned system was indeed an engine is also directly supported by the authors themselves in the conclusion of their paper, which states "*However, in this study, compared with the obtained free energy of ≈ kT, a huge amount of energy was consumed for the information processing at the macroscopic level.*" In other words, no energy was produced because running the measurement and control process required many orders of magnitude more energy than the potential energy gain ≈ kT by the particle. To experimentally demonstrate that Szilard's engine or Maxwell's demon work—or, more precisely, prove that their energy balance is negative—one would need to record heat and work on the $kT$ energy level, which is technically impossible with today's measurement techniques.

In [22] a single-bit memory was constructed by placing a particle in a double-well potential and the state of the memory was changed while the energy dissipation component by moving the particle against the viscous forces was evaluated. The flipping of the state of this single-bit memory was claimed to be an erasure process and the energy dissipation by the viscous forces seemed to satisfy Landauer's theorem. While the experiment is interesting, it cannot be the experimental proof of the theorem due to several fundamental reasons:

(i) The paper [22] defines erasure as a process that ends the particle in a given position independently from its starting position. What would change in this experimental process, if instead of erasure, the issue was writing-in a new information? Obviously, nothing would change because this is a single-bit memory. Thus erasure has no special role under these conditions, and the same energy dissipation will take place during writing new information instead of erasure. This contradicts to Landauer's picture.

(ii) Is the dissipation by viscous forces all the heat generated in this process? How about the laser light providing the potential well and its control? We are talking here about astronomically great energy dissipation compared kT ln(2). Thus the energy dissipation component evaluated in [22] is a negligible part of all the energy dissipation required to change the state of the memory consequently it cannot prove or disprove the validity of Landauer's theorem.

**7. Conclusion** – In Szilard's engine, the generated control information during repositioning of the piston demands the dissipation of at least $5kT\ln(2)$, which is five times greater than the average energy gain in a single cycle. Thus the energy requirement of control, which is an essential part of the Szilard's engine, will itself ascertain that the Second Law of Thermodynamics is not violated.

In Maxwell's demon, the lower limit of energy dissipation for controlling the trapdoor is even greater and amounts to $7kT\ln(2)$. The selection of molecules with energies higher than this value is possible at the expense of an exponential decrease of the ratio of selected molecules versus all of the molecules that have been measured. Thus when one separates the measurement process of molecules from its purpose of controlling the trapdoor, the measurement process dominates the ultimate energy requirement and again saves the Second Law of Thermodynamics.

Finally, we would like to mention some recent works aiming a full quantum mechanical description of these demons (for example [24, 25]: in the published quantum physical description of these demons, the treatment of energy dissipation of controlling the necessary classical physical objects: piston, trapdoor, or corresponding potential, is completely missing. One side of the coin has been thoroughly studied wile the other side, which turns out to be essential, has been neglected.

In the present paper, we have addressed this unexplored issue and showed that it changes the picture about the energy balance of the Szilard engine.

ACKNOWLEDGEMENTS

Critical reading and valuable feedback by John Norton are greatly appreciated. We are grateful for discussions and encouragement to Dave Ferry, Wolfgang Porod, Julio Gea-Banacloche, Gabor Balazsi, Mihaly Benedict, Janos Hajdu, Janos Kertesz and Zoltan Racz.

REFERENCES

[1] KISH L. B., *Chaos, Solitons and Fractals,* **4 4** (2011) 114-121.

[2] SZILARD L., *Z. Phys.,* **53** (1929) 840-856.

[3] MAXWELL J. C., *Theory of Heat* (Longmans, Green & Co., London) 1871.

[4] LEFF H. S., REX A. S. (Editors), *Maxwell's Demon 2: Entropy, Classical and Quantum Information, Computing* (Institute of Physics, Bristol) 2003.

[5] BRILLOUIN L., *Scientific Uncertainty and Information*, (Academic Press, New York) 1964.

[6] BRILLOUIN L., *Science and Information Theory*, (Academic Press, New York) 1962.

[7] NORTON J. D., *Stud. Hist. Philos. Mod. Phys.,* **42** (2011) 184-198.

[8] NORTON J. D., *Stud. Hist. Philos. Mod. Phys.* **36** (2005) 375-411.

[9] NORTON J. D., "The end of the thermodynamics of computation: A no go result", manuscript; http://www.pitt.edu/~jdnorton/papers/No_Go.pdf

[10] POROD W, *Appl. Phys. Lett.,* **52** (1988) 2191-2191.

[11] POROD W., GRONDIN R. O., FERRY D. K., *Phys. Rev. Lett.,* **52** (1984) 232-235.

[12] POROD W., GRONDIN R. O., FERRY D. K., POROD G., *Phys. Rev. Lett.,* **52** (1984) 1206-1206, and references therein.

[13] KISH L. B., *Phys. Lett. A,* **305** (2002) 144-149.

[14] KISH L. B., *IEE Proc.—Circ. Dev. Syst.,* **151** (2004) 190-194.

[15] KISH L. B., *Appl. Phys. Lett.,* **89** (2006) 144104.

[16] GEA-BANACLOCHE J., KISH L. B., *Proc. IEEE,* **93** (2005) 1858-1863.

[17] GEA-BANACLOCHE J., KISH L. B., *Fluct. Noise Lett.,* **3** (2003) C3-C6.

[18] FERRY, D., *J. Phys. Cond. Mat.,* **21** (2009) 474201 (6pp).

[19] KAK S., *Int. J. Theor. Phys.,* **45** (2006) 963-971.

[20] ALICKI R., *Fluct. Noise Lett.,* **6** (2006) C23-C28.

[21] TOYABE S., SAGAWA T., UEDA M., MUNEYUKI E., SANO M, *Nature Phys.,* **6** (2010) 988-992.

[22] BÉRUT A, ARAKELYAN A., PETROSYAN A., CILIBERTO S., DILLENSCHNEIDER R, LUTZ A., *Nature* **483** (2012) 187–189.

[23] JARZYNSKI C., *Phys. Rev. Lett.,* **78** (1997) 2690-2693.

[24] KIM S.W., SAGAWA T., LIBERATO S.D., UEDA M., *Phys. Rev. Lett.* **106** (2011) 070401.

[25] DONG H., XU D.Z., CAI C.Y. AND SUN C.P., *Phys. Rev. E* **83** (2011) 061108.